# Enhanced grain surface effect on magnetic properties of nanometric $La_{0.7}Ca_{0.3}MnO_3$ manganite : Evidence of surface spin freezing of manganite nanoparticles

**P. Dey[1], T. K. Nath[1a)], P. K. Manna[2] and S. M. Yusuf[2]**

[1]Department of Physics and Meteorology, Indian Institute of Technology Kharagpur, West Bengal, 721302, India

[2]Solid State Physics Division, Bhabha Atomic Research Centre, Mumbai 400 085, India

## *Abstract*

We have investigated the effect of nanometric grain size on magnetic properties of single phase, nanocrystalline, granular $La_{0.7}Ca_{0.3}MnO_3$ (LCMO) sample. We have considered core-shell structure of our LCMO nanoparticles, which can explain its magnetic properties. From the temperature dependence of field cooled (FC) and zero-field cooled (ZFC) dc magnetization (DCM), the magnetic properties could be distinguished into two regimes: a relatively high temperature regime T > 40 K where the broad maximum of ZFC curve (at $T = T_{max}$) is associated with the blocking of core particle moments, whereas the sharp maximum (at $T = T_S$) is related to the freezing of surface (shell) spins. The unusual shape of M (H) loop at T = 1.5 K, temperature dependent feature of coercive field and remanent magnetization give a strong support of surface spin freezing that are occurring at lower temperature regime (T < 40 K) in this LCMO nanoparticles. Additionally, waiting time ($t_w$) dependence of ZFC relaxation measurements at T = 50 K show weak dependence of relaxation rate [S(t)] on $t_w$ and dM/dln(t) following a logarithmic variation on time. Both of these features strongly support the high temperature regime to be associated with the blocking of core moments.

At T = 20 K, ZFC relaxation measurements indicates the existence of two different types of relaxation processes in the sample with S(t) attaining a maximum at the elapsed time very close to the wait time $t_w$ = 1000 sec, which is an unequivocal sign of glassy behavior. This age-dependent effect convincingly establish the surface spin freezing of our LCMO nanoparticles associated with a background of superparamagnetic (SPM) phase of core moments.

**Key Words**: Manganite, Nanoparticles, Core-shell Structure, Magnetic Property

**[a]Corresponding author.**
**E-mail address: tnath@phy.iitkgp.ernet.in**

# I. INTRODUCTION

It is well known that finite-size effects play a central role in physics, from the appearance of discrete energy levels in quantum dots to governing regimes of fluid flow. Recently, structural transitions driven by size, such as shape transitions of coherent precipitates [1] and magnetic phase transitions in ferroelectric nanosystems [2], have further highlighted the intriguing new physics that arises at reduced dimensionality. Currently, nanoscale magnetism provides a wealth of scientific knowledge and potentials for applications, which include magnetic recording media, ferrofluids, catalysis, magnetic refrigeration, medical diagnostic, bioprocessing, drug delivery system, miniaturized magnetic sensor applications, etc [3]. When the size of the magnetic particles is reduced to a few nanometers, they exhibit a number of outstanding physical properties such as giant magneto-resistance, surface spin-glass behavior, superparamagnetism, large coercivities, low-field saturation magnetization, low Curie temperature, low saturation magnetization etc., as compared to their bulk counterparts [4 – 9]. Due to the realization of these outstanding physical properties upon size reduction, magnetic nanoparticles are bringing revolutionary changes in a variety of applications. It is generally believed that a high value of the surface to volume ratio with large fraction of atoms residing at the grain boundaries is what differentiates them from the bulk materials in their properties [10] and the net magnetic behavior is dominated by surface magnetic properties [11, 12]. As for example, in case of magnetic nanoparticles the most controversial issue that is the observed reduction of the saturation magnetization has been afterwards interpreted in terms of random canting of the particles surface spins caused by competing antiferromagnetic exchange interactions at the surface as proposed by Coey [13].

Magnetic properties of manganites at the nanometer scale comprise an issue of great interest now-a-days. A number of investigations of the grain size effect on magnetic properties of perovskites $La_{1-x}A_xMnO_3$ nanoparticles have been published recently (Refs. 14 – 26). In particular, with regard to colossal magneto-resistive properties of manganites, spin electronics based on half-metallic properties of these materials, the surface spin and structure disorders and their possible influence on the magnetism of manganites nanoparticles are matters of intense discussion. From previous studies, the grain-boundary magnetic structure of manganites nanoparticles as emerged at present are as follows: since the double exchange mechanism is sensitive to Mn-O-Mn bond, any structural disorder near the grain boundary (oxygen non-stoichiometry, vacancies, stress, etc.) modifies this exchange and leads to a spin disorder. Due to a strong Hund's interaction, spin disorder around grain boundaries serves as a strong scattering center for highly spin-polarized conduction electrons and results in a high zero-field electrical resistance. The application of a moderate magnetic field can align the originally disordered Mn spins, thus reducing the scattering and leading to a giant magneto-resistance. However, as already mentioned the microscopic nature of the surface region is not well understood so far. This lack of understanding is manifested in some inconsistency between the models and the experimental results. For example, it has been assumed that the grain surface magnetization is suppressed compared to the bulk magnetization; [15, 17] meanwhile, the Curie temperature near the grain boundary was found to be enhanced; [20] some authors found a shift of the chemical potential between the external region and the inner part of the grain; [19, 23] others suggest a high probability of tunneling through paramagnetic impurity states in the intergranular barrier, [17, 18, 22] etc. Some salient features observed as we reduce the particle size of manganites systems are (a) a decrease and broadening of the ferromagnetic transition temperature $T_C$, (b) a decrease in the magnetization in comparison with single-crystal and bulk polycrystalline samples, showing superparamagnetic behavior at very low particle size. These observations can be logically explained by assuming the increase of an insulating grain-boundary contribution as the particle size decreases [4, 27 – 29]. Formation of grain boundaries causes broken bonds at the surface, which causes a decrease in the magnetization value. This is the most general observation in the case of nanoparticles of the manganite system [27, 30]. There are a few exceptions to this general rule. A recent report by Fu [31] on the $La_{0.8}Ca_{0.2}MnO_3$ nanoparticles system shows results contradicting the above facts on reduced particle size. He has shown that as the particle size decreases, the $T_C$ increases and the resistivity decreases. The decrease in the magnetization with an increase in the particle size and also the increase in the resistivity with the particle size cannot be due to the difference in the oxygen stoichiometry as explained by Fu [31]. The difference has been explained as being due to the strain at the grain boundaries. The deviation from the general rule mentioned above seems to be observed with the Sr-doped system as well, especially at low doping. Zhang *et al.* analyzed in detail the effect of the annealing temperature on the magnetization for various *x* values in the $La_{1-x}Sr_xMnO_3$ system.[32] They found that at low doping ($x < 0.25$), the magnetization decreases with an increase in the sintering temperature and for higher doping ($x > 0.25$), the magnetization increases with an increase in the sintering temperature. The lattice distortions are mainly responsible for the change of magnetization in this system according to their analysis. In the low doped system of $La_{0.85}Sr_{0.15}MnO_3$, they found that the bond angle decreased and the bond length increased as the sintering temperature increased, which explains the decrease in the magnetization they observed. The recognition and elucidation of the grain size effect is crucial if manganites are expected to be used in forthcoming nano-electronic devices. In fact, for future applications, like the magnetic recording industry, a smaller grain size of manganites system will be required.

In our previous paper [33], we presented experimental results and discussed the effect of nanometric grain size on magneto- and electronic-transport properties of single-phase, nanocrystalline, granular $La_{0.7}Ca_{0.3}MnO_3$ samples having average grain size ($\Phi$ ) in the nanometric regime (14 – 27 nm). Based upon spin-polarized tunneling mechanism and considering core-shell structure of manganites nanoparticles, we have proposed a phenomenological model to explain electronic transport behavior of mangnaites nanoparticles. Most interestingly, magneto-transport measurements on $\Phi$ = 17 and 27 nm LCMO sample showed that the magnitude of low-field magneto-resistance, as well as of high - field magneto-resistance remains constant up to sufficiently high temperature (~ 220 K) and then drops sharply with temperature. The effect gets more pronounced with the decrease in particle size. Analyzing our data following the theoretical perspective of S. Lee *et al.* [34], we found that this strange temperature dependence of MR is decided predominantly by the nature of the temperature response of surface magnetization of nanosize magnetic particles. In order to further shed light on this issue and to investigate the surface magnetic properties of our LCMO nanoparticles, we have further carried out magnetization studies on LCMO sample having smallest grain size ($\Phi$ = 17 nm) of our series. For this purpose, we have carried out dc magnetization (M) and zero-field-cooled (ZFC) relaxation measurements in the temperature range of 1.5 – 300 K and in the magnetic field ($H$) range of – 5.5 - 0 - + 5.5 T. Similar to electronic- and magneto-transport properties, all of our magnetic results can be coherently analyzed considering core-shell structure of this LCMO nanoparticles. From the temperature dependence of field cooled (FC) and ZFC dc magnetization, the magnetic properties could be distinguished into two regimes: a relatively high temperature regime $T \geq 40$ K where the broad maximum of ZFC curve (at $T = T_{max}$) is associated with the blocking of core particle moments, whereas the sharp maximum (at $T = T_S$) is related to the freezing of surface (shell) spins. In fact, in a previous Letter [4] Zhu *et al.* reported similar feature of FC - ZFC magnetization, which they have attributed to the surface spin glass behavior. Furthermore, shape of M ($H$) loop at T = 1.5 K, temperature dependent feature of coercive field and remanent magnetization and the waiting time dependence of ZFC relaxation measurements at T = 50 and 20 K gives strong support of surface spin freezing occurring at lower temperature regime (T < 40 K) in this LCMO nanoparticles, associated with a background of SPM phase of core moments. For comparison, we have also carried out both magnetic field and temperature dependence of FC and ZFC dc magnetization measurements for a large grain size (Φ ~ 27 nm) sample. Although, our magnetization study give evidences of surface spin freezing for this Φ ~ 27 nm LCMO sample as well, the effect gets reduced, which strongly establishes enhanced grain surface on magnetic properties with decreasing grain size of manganite nanoparticles.

## II. EXPERIMENTAL DETAILS

Nanometric particles of $La_{0.7}Ca_{0.3}MnO_3$ (LCMO) were prepared from high-purity $La_2O_3$ (99.99 %), $Mn(CH_3COO)_2$ (99 %) and $CaCO_3$ (99 %) by “pyrophoric reaction process". The advantages of chemical synthesis technique, involving liquid solutions, in preparing oxide nanoparticles is that it promises high purity, small particles sizes with good particle-size distributions, good compositional control and chemical homogeneity of the final products. Details of state of the art in the synthesis of oxide powders through pyrophoric reaction process are already published in the literature by Pathak *et al.* [35]. This pyrophoric reaction process is a novel and versatile chemical technique that has been developed for the preparation of nanosized ceramic powders, using thermolysis of aqueous precursor solutions of coordinated metal compounds of organic amines and acids via the formation of mesoporous carbon precursors and their calcination at temperatures < 800 K. The principle is to atomistically disperse the complexed metal ions in the polymeric network provided by organic coordinating agent, i.e., triethanolamine (TEA), during pyrolysis of excess reagents. During pyrolysis of the precursor solution, the metal ions or their salts form nanoclusters, which are embedded in the resulting matrix of mesoporous carbon. Slow volatilization of mesoporous carbon in the precursor material through low-temperature (between 500 and 800 K) air oxidation, aided by the catalytic effect of in-situ metal ions, favours the formation of metal-oxide nanocrystals obtained at relatively lower pyrolysis temperatures than those reported to date in the literature.

The chemical reactions involved in this method are as follows:

$$0.7\ La(NO_3)_3 + 0.3\ CaCO_3 + N(CH_2CH_2OH) \rightarrow [La - N(CH_2CH_2OH)_3]^{2+} +$$

$$[Ca - N(CH_2CH_2OH)_3]^{2+} + [Mn - N(CH_2CH_2OH)_3]^{2+} + NO_3^-$$

$$\rightarrow \text{oxidation in air}$$

$$\rightarrow La_{0.7}Ca_{0.3}MnO_3\ \text{(nanosized)} + CO_2 + NO_2 + N_2 + H_20$$

The dried carbonaceous mass is then ground to fine powder and have been calcined at various temperatures to get a series of LCMO nano crystalline powders. The heat treatments of the precursor materials (in air 5 h) have been facilitated from 600 to 1000ºC.

Structural characterizations of LCMO nanoparticles were carried out using x-ray powder diffraction (XRD), transmission electron microscopy (TEM), high resolution field emission scanning electron microscopy (FE-SEM) and energy dispersive x-ray spectroscopy (EDAX). XRD (Models PW 1710 and PW 1810, Philips) was performed with monochromatic Cu-$K_{\alpha}$ radiation ($\lambda \sim 1.542$ Å). TEM and EDAX were carried out employing a JEOL 2100F UHR version electron microscope (equipped with an EDAX unit) at an accelerating voltage of 200 kV. Specimens for TEM and EDAX were prepared by dispersing samples in acetone in suitable concentrations using an ultrasonic device and were dropped to the amorphous carbon coated copper grids. After being well dried in air, the grid was mounted on the TEM specimen holder for examination. FE-SEM measurements were carried out using Carl Zeiss SMT Ltd. SUPRA™ 40. Dc magnetization measurements were carried out as a function of temperature and magnetic field using a commercial vibrating sample magnetometer (Oxford Instruments). The field dependence of magnetization, at various temperatures, was carried out after cooling the sample from room temperature down to the measurement temperature in absence of field. For zero-field-cooled (ZFC) magnetization, the sample was first cooled from room temperature down to 5 K in zero field. After applying the magnetic field at 5 K, the magnetization was measured in the warming up cycle with field on. Whereas, for field-cooled (FC) magnetization measurements, the sample was cooled in the same field (measuring field in the ZFC case) down to 5 K and FC magnetization was measured in the warming up cycle under the same field. Temperature dependence of the remanent magnetization was measured in the following way : first the sample was cooled under a field of 50 kOe from room temperature down to 5 K, then the field was switched off and thermo remanent magnetization (TRM) was measured heating up to room temperature. ZFC magnetic relaxation measurements were performed at T = 20 and 50 K. The measurements were performed as follows : first the sample was cooled in zero field down to the measuring temperature ($T_m$); at $T_m$, 10 sec elapsed (waiting time, $t_w$) before the application of the magnetic field ($H$ = 50 Oe) and then the time variation of the ZFC magnetization was recorded. At the end of this first measurement, the field was removed and the temperature was raised to 300 K, lowered again at the same $T_m$ as before and, after $t_w$ = 1000 sec, the magnetic field was applied and the time variation of ZFC magnetization was recorded again.

## III. RESULTS AND DISCUSSIONS

### a) Structural Characterization of the samples

The phase purity and crystal structure of our LCMO nanoparticles, having two different grain sizes, were checked by XRD at room temperature. Figure 1 (a) shows room-temperature XRD pattern of LCMO samples, obtained by calcining the precursors at different calcination temperatures, $T_{cal}$ = 650 and $800^0$ C. XRD profiles [Fig. 1 (a)] confirm the pure single phase of the samples without any secondary phase within the detection limit of our instrument and that the samples have cubic perovskite structure showing the characteristic peaks of the perovskite. All the peaks were indexed on the basis of cubic perovskite phase.

XRD lines of these powders are very broad with large FWHM [inset in Fig. 1 (b)], indicating the formation of nanocrystalline fine LCMO powders. The average grain size ($\Phi$), calculated using Scherrer formula: $\Phi = k\lambda/\beta_{eff} \cos\theta$, where $\Phi$ is the diameter of the nanocrystals in Å, k is shape factor (generally taken as 0.89), $\lambda$ is the wavelength of Cu $K_{\alpha}$ radiation (1.542 Å), $\theta$ is the diffraction angle and $\beta_{eff}$ is defined as $\beta_{eff}^2 = \beta_m^2 - \beta_s^2$, where $\beta_m$ and $\beta_s$ are the experimental FWHM of the present sample and that of a standard silicon sample, respectively. The standard silicon sample was used to calibrate the intrinsic width associated with the equipment. We get $\Phi$ = 17 and 27 nm corresponding to the calcination temperature of $T_{cal}$ = 650 and $800^0$ C, respectively. It is evident from Fig. 1 (b) that increase in the calcination temperature from $T_{Cal}$ = 650 to 1000°C resulting in increase in average $\Phi$. We have calculated the values of $\Phi$s from different peaks of the XRD pattern, corresponding to different (*h k l*) planes, for our series of LCMO samples. The distribution bars to the average crystal sizes, as shown in Fig. 1 (b), indicate increasing size dispersion of our samples with increase in $T_{Cal}$s. This feature mimics the grain size dispersion feature of LCMO nanoparticles from previous literatures [36].

Structural characterization through TEM provides visual demonstration to estimate grain size exactly. We have carried out TEM studies on our $\Phi$ = 17 nm LCMO sample, calcined at 650° C. Figures 2 show typical TEM images, selected area electron diffraction (SAED) pattern and high - resolution TEM (HRTEM) of LCMO nanocrystals. Bright field TEM images [Fig. 2 (a) and inset in 2 (b)] of $\Phi$ = 17 nm LCMO sample indicate that the sample is dispersive single crystal particles having polyhedron shapes with hexagonal projections [37]. The images show an abundance of particles whose size distribution is given by the histogram shown in inset in Fig. 2 (b), the histogram being obtained by analyzing several frames of similar bright field images using Image J software. We find that the particles have an average size of ~ 17 nm, which is in close agreement with the results obtained from XRD studies (~ 17 nm). Selected area electron diffraction pattern (SAED) [Fig. 2 (b)] recorded on many nanoparticles indicates that they are crystalline in nature. From the reflection spots, we have estimated the interplanner lattice spacing *d* using the formula, $\lambda L = d D$, where λ = wavelength of electron beam, $L$ = effective camera length resulting from the magnifications of the imaging lenses of the microscope column, D = the distance on the diffraction pattern from the origin to a diffracted spot. In fact, the distances measured on the diffraction pattern are actually magnified reciprocal lattice vectors. The interplanner lattice spacing '$d$' is found out to be ~ 2.830 (± 0.001) Å, which corresponds to the (110) plane of the crystal within our accuracy of estimating D. This result, in turn, justifies the XRD result for our Φ = 17 nm LCMO sample, where (110) plane correspond to the average crystal size of ~ 17 nm.

We have carried out FE-SEM study on our Φ = 17 nm LCMO sample. The measurement was performed on the pellet of the sample. Figures 3 show typical FE-SEM cross-section micrographs for Φ = 17 nm LCMO sample, which exhibits an abundance of nearly spherical particles in the frame of FE-SEM micrographs. We find that the particles have an average size of ~ 20 nm, which is in close agreement with the results obtained from XRD studies (~ 17 nm) as well as from TEM micrographs (~ 17 nm). For compositional analysis, we have carried out EDAX for Φ = 17 nm LCMO sample of several particles. Figure 4 exhibits EDAX spectrum for Φ = 17 nm LCMO nanoparticles at various locations. The EDAX spectrum does not vary appreciably at various locations in a single sample and appears to be identical within an experimental error, thus confirming good chemical homogeneity of the sample. Furthermore, from the weight % of different constituent elements we have estimated the formula units of the sample, which is found out to be $La_{0.69}Ca_{0.35}Mn_{1.12}O_{2.99}$. Evidently, this estimated formula unit is nearly identical with the *desired* stoichiometry of the nanocrystals, based on which we prepared this sample and also confirm quantitatively good chemical homogeneity of our sample.

## b) Magnetic Property of the samples

We have investigated field-cooled and zero field-cooled temperature dependence of dc magnetization (DCM) at a magnetic field of H = 50 Oe for Φ = 17 nm LCMO nanoparticles [Fig. 5]. DCM measurements as a function of temperature were performed according to the standard zero-field cooling (ZFC) and field cooling (FC) procedures. These measurements exhibit strong irreversibility between FC and ZFC curves as indicated by the appearance of large bifurcation between them. As is evident from Fig. 5, ZFC curves exhibit a rather broad peak at $T_{max}$, (i.e., the temperature of the maximum in ZFC) and the FC curves continue to increase with decreasing temperature. This strong history dependence is generic feature of several commonly known metastable magnetic systems like spin glasses, cluster glasses and superparamagnets and was also seen in randomly canted ferromagnets with perovskite structures [38, 39]. However, the features that the two curves depart from each other at much higher temperature (designated as $T_{irr}$) than $T_{max}$ along with the rather broad ZFC peak and the FC magnetization continues to increase without saturation below $T_{max}$, distinguish this system from spin glass systems and hints towards superparamagnetic (SPM) phase associated with this system [40]. It is evident from Fig. 5 that FC – ZFC DCM exhibits an irreversibility on the magnetization below $T_{irr}$, which is typical of SPM single-domain particles characterized by a regime at high temperatures and the blocked regime at T < $T_{max}$. In fact, the broad peak at ZFC curve at $T_{max}$ marks a crossover region where the average anisotropy energy and the energy caused by the thermal energy ($k_B T$) are comparable and the anisotropy energy is predominant for T < $T_{max}$ and so is the later at T > $T_{max}$. The observed behavior reveals the progressive blocking of the SPM particle moments, with a distribution of relaxation times related to the size and anisotropy axis direction distributions [40]. In general, for an assembly of non-identical, non-interacting magnetic nanoparticles, the low field M(T) depends on the type of anisotropy energy barrier distribution function, which governs the relationship between $T_{max}$ and the average blocking temperature $<T_B>$. For a particle of volume *V*, $T_B$ is defined as the temperature at which the relaxation time, described by the Néel-Brown expression, $\tau = \tau_0 \exp(E_B / k_B T)$, where $E_B$ = KV, becomes equal to the measuring time $t_m$. $T_{irr}$ corresponds to the highest blocking temperature, i.e., to that of particles with highest energy barrier.

Interestingly, at very low temperatures a second sharp maximum in $M_{ZFC}$ is observed for both H = 50 [Fig. 6 (b)] and 100 Oe at $T_S$, associated with a strong decrease of low-field ZFC DCM with further decreasing temperature below $T_S$. Additionally, there is sharp rise of FC curve at that temperature of $T_S$ [Fig. 6 (a)]. This sharp rise in FC curve is followed by a flattening of FC magnetization with further lowering temperature below $T_S$. Similar magnetic results were also reported by other groups for manganites nanoparticles [4], as well as for dispersed amorphous nanoparticles [41, 42]. We have tried to interpret our results within the core-shell model of our LCMO nanoparticles. In our previous paper [33], in order to describe the electrical - and magneto - transport properties of this same series of nanometric LCMO samples, we have considered core-shell structure of the nanoparticles. Practically, when the size of the manganites grain reduces to few tens of nanometers, we can assign a core-shell structure to them [4, 14, 27]. In fact, due to the nanometric grain size of our manganites samples, the surface-to-volume ratio is sufficiently large and, as a result, the following physical effects are most likely to take place in a higher degree. Those are (a) contamination of the grain surface, (b) breaking of Mn-O-Mn paths at the grain surface, (c) deviation of stoichiometric composition at the grain surface, (d) termination of the crystal structure at the grain surface and (e) dislocation at the grain boundaries. As a result, the inner part of the grain, i.e., the core, would have the same properties as the bulk compound, whereas the outer layer, i.e., shell would contain most of the oxygen defects and faults in the crystallographic structure. Again, at the manganites grain surface, there may be antiferromagnetic ordering of the Mn spins due to the modification of charge state of the Mn ions [43]. This combination of topological disorder and competing magnetic interactions may result in a magnetically disordered state at the grain surface since the links holding the surface spins, aligned by the core of the ferromagnetic grains, are progressively severed. Consequently, competing magnetic interactions stabilize a spin-glass like state at the surface region of the grain, yielding freezing of those surface spins in random directions. This, in turn, inhibits the exchange interaction to transmit across the interfaces resulting in uncoupled assembly of nanoparticles.

Within the consideration of core-shell model, the broad maximum of ZFC at $T_{max}$ is associated with the blocking of core particle moments, whereas the sharp maximum at $T_S$ is related to the freezing of surface (shell) spins. With decreasing temperature, surface spin fluctuations slow down and short range correlation among them develop progressively in magnetic correlated spins regions of growing size. The flattening of the FC curve after sharp rise below $T_S$ and the negligible field dependence of $T_S$ (not shown here) corroborates our understanding of this anomaly arising at $T_S$ associated with any kind of spin-glass like transition. In this case, the FC curve does not superimpose the ZFC curve at temperatures above $T_S$ because the SPM component associated with this behavior has an additional background due to the FC magnetization of the blocking at higher temperature.

We have also performed isothermal magnetization measurements as a function of magnetic field at different temperatures after ZFC as shown in Fig. 7. At T = 200 K, i.e., at the vicinity of $T_{irr}$, M (H) curve does not present hysteresis as expected in the unblocked regime [upper inset in Fig. 7(b)]. For $T < T_{max}$ [Figs. 7 (b) – 7 (f)], hysteresis behavior occurs in M (H) curves, which is quite expected in the blocked regime of any SPM system. Interestingly, even for $T_{max} < T < T_{irr}$, i.e., at T = 80 and 150 K as shown in Fig. 7 (a), we observe hysteresis behavior in M(H) curves, which we attribute to the growing fraction of blocked particles moments with decreasing temperature. This, in turn emboldens the broad particle size dispersion of our LCMO nanoparticles. As is shown in lower inset of Fig. 7 (b), the magnetization of the sample at T (= 200 K) > $T_{irr}$ (i.e., in the unblocked regime) clearly consists of two components, $M(H) = M_{SP}(H) + \chi H$, where $M_{SP}$ is a SPM one which follows Langevin-like function and $\chi H$ is an extra paramagnetic term. We believe the physical significance of the inclusion of this extra paramagnetic term lies on the approximately linear response of surface spins to applied magnetic field, which in turn indicates a paramagnetic contribution of surface spins at this higher (220 K) temperature [43]. Assumption of the distribution of pinning energy barrier of surface spins (residing at the shell portion) of our nanoparticles from practical considerations can take account this non-saturating behavior of M (H) curves up to a high magnetic field of 5.5 T, described approximately by an extra paramagnetic term ($\chi$ H). In general, isothermal magnetization curves as a function of field of this present sample display two typical features of any fine particle systems: (a) M (H) curve show hysteretic behavior without saturation of the magnetization [Fig. 7 (a) – 7 (f)], (b) the magnetization value at our maximum field of 5.5 T is still about half of the bulk saturation value of 3.7 $\mu_B$/ formula unit.

Noticeably, a change of hysteretic loop shape is observed below and in the vicinity of $T_S$ (~ 40 K) for T = 1.5, 10 and 25 K [Figs. 7 (b), (c) and (d)], characterized by a narrowing of cycle at low field regime where the demagnetization and remagnetization curves vary rapidly and irreversibility between demagnetization and remagentization curves persist up to a sufficient large magnetic field. Moreover, both remanent magnetization ($M_r$) and coercive field ($H_C$), as obtained from each isothermal M (H) curves, exhibit rapid increase below $T_S$ (~ 40 K) [Fig. 8 and its inset], i.e., there is a distinct change in the slope of both of these curves at $T_S$. Furthermore, we have also studied the temperature dependence of the remanent magnetization $M_r$ (T), i.e., TRM. This TRM curve also exhibits a rapid decrease above $T_S$. Thus, M (H) curves [as shown in Fig. 7], Fig. 8 and TRM curve [as shown in Fig. 9 (a)] suggest the existence of two contributions to the magnetization of the present sample. Within the understanding of core-shell structure of our LCMO nanoparticles, the high temperature behavior is determined by the core contribution and is weekly temperature dependent, whereas at low temperature the large magnetization is determined by the shell contribution. In fact, the magnetization in the present sample comes for the superposition of two contributions: one from the spins of the ferromagnetic particle core, which tends to saturate at low fields and the other from the magnetically disordered surface frozen spins, which does not saturate even up to a field of 50 kOe [from the hysteresis loops for $T \leq T_S$, Fig. 7 (b) – 7 (f)]. The later is thus responsible for the change in the shape of hysteresis loop at T close to $T_S$ and for the change in the temperature dependence of $H_c$ and $M_r$ for $T \leq T_S$, which has also support from previous literatures [41, 42].

Furthermore, above T = 45 K, $H_c$ varies as $T^{1/2}$ according to the Stoner-Wohlfarth model [44] for an assembly of particle moments with uniaxial anisotropy and a random distribution of anisotropy axes, where $H_C$ can be expressed in terms of temperatures and blocking temperature $T_B$ of SPM particles [45] as,

$$H_C = H_{C,0}[(1 - T/T_B)^{1/2}] \qquad (1)$$

where, $H_{C,0}\left(= \frac{2K}{M_S}\right)$ is $H_C$ when T approaches zero and $T_B$ will have single value when the assembly of SPM particles are of constant size, whereas for broad distribution of particle sizes $T_B$ is associated either with the maximum or average of some large $T_B$s of the distribution. Inset in Fig. 9 (a) shows the plot of $H_C$ as a function of $\sqrt{T}$, where straight line is the linear fit of experimental coercivity data to Eq. 1 exhibiting a quite satisfactory fit. The value of $T_B$ and $2K/M_S$ as extracted from this fit is ~ 204 K and 321, respectively. As shown in inset of Fig. 9 (a), LCMO nanoparticles closely follow Eq. 1 for $T \geq 45$ K, thus supporting that above this temperature the magnetic behavior is governed by the SPM phase of the particle core contribution to the magnetization. We attribute the deviations from Eq. 1 of $H_C$ versus $\sqrt{T}$ curve below 45 K to the progressive blocking of surface spin correlated region effective moments. Furthermore, M-H curve of this sample at T = 200 K, i.e., in the vicinity of that estimated $T_B$ of 204 K, shows the typical characteristics of SPM behavior exhibiting almost immeasurable coercivity and remanance [45] [inset in Fig. 8 (a)]. These features of M-H curve in fact provide strong support of SPM phase associated with this nanoparticle system [45]. Moreover, these features also confirm the absence of any FM contribution, arising from large particles of the distribution. In addition, observation of those two above said typical characteristics of SPM behavior even at T = 200 K, further suggests a reasonably good estimation of maximum/upper limit of $T_B$ (~ 204 K) of the distribution from fit of Eq. 1.

In order to further probe the nature of low temperature phase transition at $T_S$ of this sample, we have performed aging experiments at T = 20 (not shown) and 50 K [Fig. 9 (b)] for two different wait times $t_w$ = 10 and 1000 sec (registered as described in the experimental section). In general, the aging effect is observed in disordered spin systems governed by correlated dynamics and is related to the evolution of the spin-glass system within the characteristic multivalley free energy landscape. The former temperature is below $T_S$, whereas the later is below $T_{max}$ but above $T_S$, i.e., for $T_S \leq T \leq T_{max}$. The magnetization (M) versus time (t) curves, as shown in Fig. 9 (b), are clearly dependent on the elapsed time $t_w$. However, $t_w$ dependence of the relaxation is weaker at T = 50 K than that of T = 20 K. Another feature is the decrease in magnetization with increase in $t_w$. Relaxation rate of the time-dependent relaxation of the ZFC magnetization M (t) is defined as [46],

$$S(t) = \frac{1}{h}\frac{dM(t)}{d\ln t} \quad \text{.........................................................................(2).}$$

Inset in Fig. 10 (a) shows S (t) at T = 50 K for $t_w$ = 10 and 1000 secs. It appears that S (t) is found to weakly dependent on $t_w$. Moreover, the time variation of the logarithmic time derivative of the zero-field-cooled magnetization i.e., $dM/d\ln(t)$ is found to follow a logarithmic variation on time [Fig. 10 (a)]. It is well established that individual particle relaxation should follow logarithmic variation with no influence of the waiting time. Therefore, the observation of both of these features in our sample support SPM blocked state associated with this temperature T = 50 K [47]. Thus it is now conclusive to assert that $T_{max}$ is associated [Fig. 6] with the blocking of core moments of our LCMO nanoparticles within the consideration of core-shell model.

Figures 10 (b) and (c) show the time variation of the logarithmic time derivative of the zero-field-cooled magnetization i.e., $dM/d\ln(t)$ at T = 20 K for $t_w$ = 10 and 1000 sec, respectively. It is evident from $dM/d\ln(t)$ versus ln ($t$) curves, as shown in Figs. 10 (b) and (c), that two different types of relaxation processes are present in the sample, i.e., there are mixing of two phases at this temperature. This seems to embolden our understanding of the sharp maximum at $T_S$ [Fig. 6 (b) and (d)] to be related with low temperature freezing of surface (shell) spins with the SPM component of core moments, exhibiting broad maximum at $T_{max}$, associated as an additional background. Furthermore, inset in Fig. 10 (c) exemplifies relaxation rate, S($t$) for $t_w$ = 1000 sec at T = 20 K. It is clear that S (t) attains a maximum at the elapsed time very close to the wait time $t_w$ = 1000 sec indicating convincingly an age-dependent effect, which is often observed in spin glass, and even in the ferromagnetic phase of a reentrant spin glass and the two dimensional random-exchange Ising ferromagnet [46, 48]. Thus this observation [inset in Fig. 10 (c)] is conclusive in asserting chaotic correlated dynamics at the surface region of our LCMO nanoparticles associated with a background of SPM phase of core moments.

## IV. CONCLUSIONS

In conclusions, we have presented detailed studies of microstructural and magnetic properties of LCMO sample having a grain size of few tens of nanometer. Room temperature XRD profiles confirm the pure single cubic perovskite phase with large full width at half maximum indicating the formation of nanocrystalline (~ 17 nm) fine LCMO powders. Both bright field TEM and FE-SEM image confirm the nanometric particle size (~ 20 nm) of our samples and exhibit polydisperse Φs. EDAX spectra confirm quantitatively the homogeneity and also almost identical chemical composition with our *desired* stoichiometry of LCMO nanoparticles. Dc magnetization and zero-field-cooled (ZFC) relaxation measurements in the temperature range of 1.5 – 300 K and in the magnetic field range of – 5.5 - 0 - + 5.5 Tesla are carried out on LCMO nanoparticles. We have considered core-shell like structure in these LCMO nanoparticles, which enable us to coherently analyze all those experimental results. From the temperature dependence of field cooled (FC) and ZFC dc magnetization, the magnetic properties could be distinguished into two regimes: a relatively high temperature regime $T \geq 40$ K where the broad maximum of ZFC curve (at $T = T_{max}$ ) is associated with the blocking of core particle moments, whereas the sharp maximum (at $T = T_S$) is related to the freezing of surface (shell) spins. We have observed an unusual shape of low temperature isothermal ($T = 1.5$ K) magnetic field dependent magnetization M (H) measurements that can be understood in terms of surface spin freezing of LCMO nanoparticles. Additionally, the temperature dependent feature of coercive field and remanent magnetization ($M_r$) gives strong support of surface spin freezing. Our ZFC relaxation measurements of magnetization for waiting times, $t_w$ = 10 and 1000 sec at $T = 50$ K, show that the relaxation rate S (t) is weakly dependent on $t_w$. Moreover, the time variation of the logarithmic time derivative of the zero-field-cooled magnetization i.e., $dM/d\ln(t)$ is found to follow a logarithmic variation on time. These two features strongly support that the high temperature regime (T > 40 K) is associated with the blocking of core moments of our LCMO nanoparticles. On the other hand, ZFC relaxation measurements at T = 20 K indicates the existence of two different types of relaxation processes in the sample. Importantly, S (t) attains a maximum at the elapsed time very close to the wait time $t_w$ = 1000 sec, which is an unequivocal sign of glassy behavior. Thus this age-dependent effect convincingly point out surface spin freezing of our LCMO nanoparticles associated with a background of SPM phase of core moments.

# Figure Captions :

**Figure 1** (Color online) **(a)** XRD patterns of LCMO samples calcined at $T_{Cal}$ = 650 and 800° C. All the peaks are indexed considering pseudo cubic perovskite phase. **(b)** Variation of Φ s, obtained from Scherrer formula, with $T_{Cal}$s. Inset shows the Gaussian fit of most intense peak for Φ = 17 nm LCMO sample, having $T_{Cal}$ = 650° C.

**Figure 2** **(a)** TEM micrograph for LCMO sample having Φ = 17 nm. **(b)** SAED pattern of LCMO nanocrystals showing the single crystalline nature of LCMO nanograins. Inset shows bright field TEM images of LCMO nanocrystals. Inset shows histogram of the grain size distribution.

**Figure 3** FE-SEM images for LCMO nanoparticles calcined at $650^0$ C (Φ = 17 nm).

**Figure 4** (Color online) EDAX spectra for LCMO nanoparticles, calcined at $650^0$ C (Φ = 17 nm).

**Figure 5** (Color online) Field cooled (FC) and zero-field cooled (ZFC) dc magnetization (DCM) as a function of temperature at H = 50 Oe for Φ = 17 nm LCMO nanoparticles in the temperature range of 4 – 300 K.

**Figure 6** (Color online) **(a)** FC and **(b)** ZFC DCM at a narrow range of temperature from 4 – 85 K at H = 50 Oe for Φ = 17 nm LCMO nanoparticles.

**Figure 7** (Color online) **(a)** M (H) hysteresis loops at different temperatures from T = 1.5 to 300 K in the magnetic field range of – 5.5 - 0 - + 5.5 T for Φ = 17 nm LCMO nanoparticles. Inset shows the same plots at a narrow range of magnetic field of – 2 - 0 - + 2 T. **(b)** M (H) hysteresis loop at T = 1.5 K in the magnetic field range of – 5.5 - 0 - + 5.5 T showing a change in the shape of the hysteresis loop. Upper inset shows M (H) hysteresis loop at T = 200 K at a narrow range of magnetic field of – 1 - 0 - + 1 T. Lower inset shows the first quadrant of the same plot at T = 200 K up to a magnetic field of 5.5 T. M (H) hysteresis loops in the magnetic field range of – 2 - 0 - + 2 T at different temperatures of T = **(c)** 10, **(d)** 25, **(e)** 45 and **(f)** 60 K.

**Figure 8** (Color online) $H_C$ as a function of temperature for Φ = 17 nm LCMO nanoparticles. Inset shows remanent magnetization ($M_r$) as a function of temperature.

**Figure 9** (Color online) **(a)** TRM as a function of temperature for Φ = 17 nm LCMO nanoparticles. Inset shows $H_C$ as a function of $\sqrt{T}$ for the same sample, where symbols are the experimental data and red solid line is the fitting to Eq. 1. Time dependence of the ZFC magnetization, measured at **(b)** $T_m$ = 50 K for two different waiting times (10 and $10^3$ sec) for Φ = 17 nm LCMO nanoparticles.

**Figure 10** (Color online) **(a)** Time variation of the logarithmic time derivative of the zero-field-cooled magnetization i.e., $dM / d\ln(t)$ at 50 Oe for **(a)** T = 50 K and waiting time $t_w$ = 10 and 1000 sec, **(b)** for T = 20 K and waiting time $t_w$ = 10 sec and **(c)** for T = 20 K and waiting time $t_w$ = 1000 sec, for Φ = 17 nm LCMO nanoparticles. Inset in **(a)** and **(c)** shows the corresponding relaxation rate S (*t*).

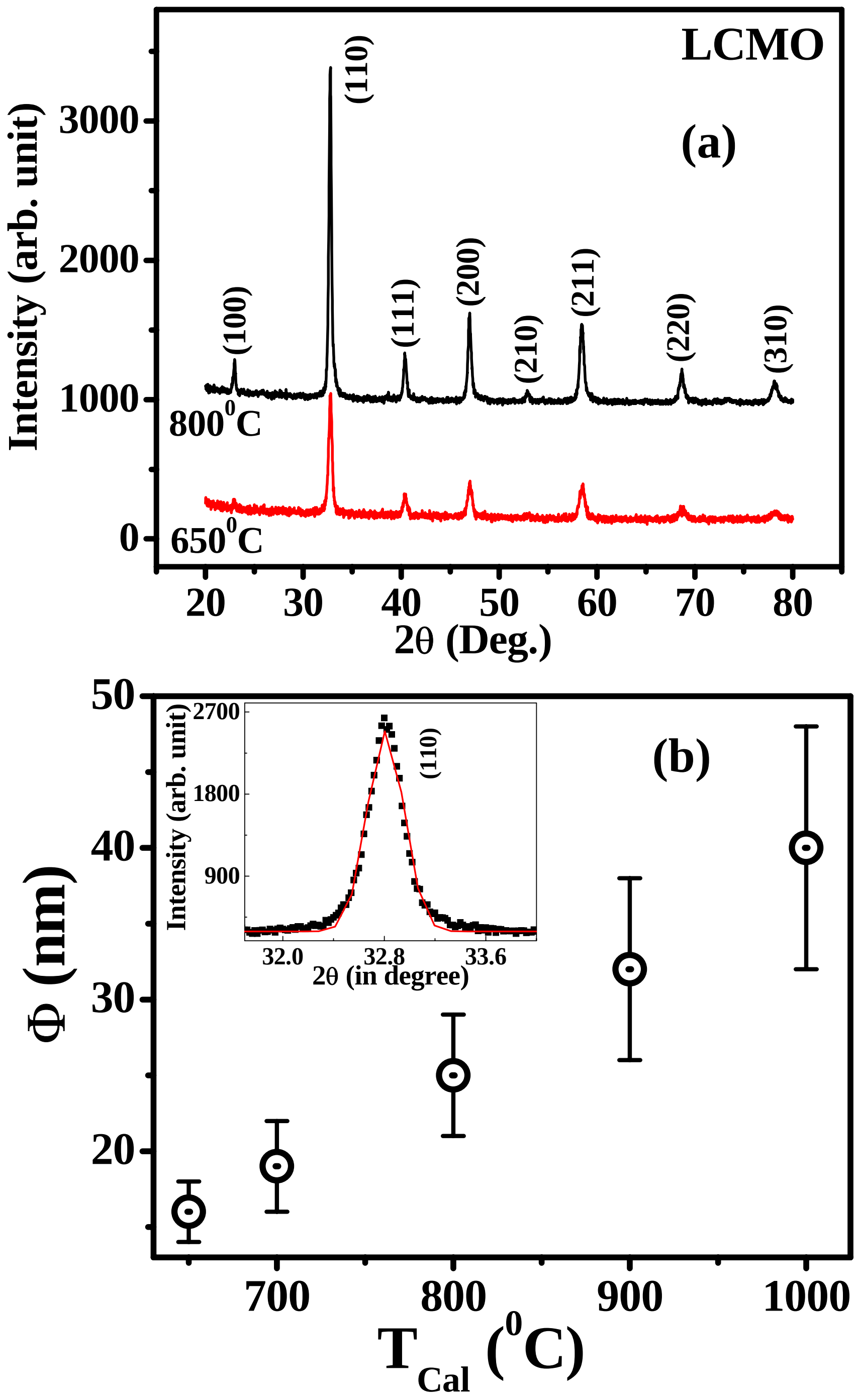

**Fig. 1**

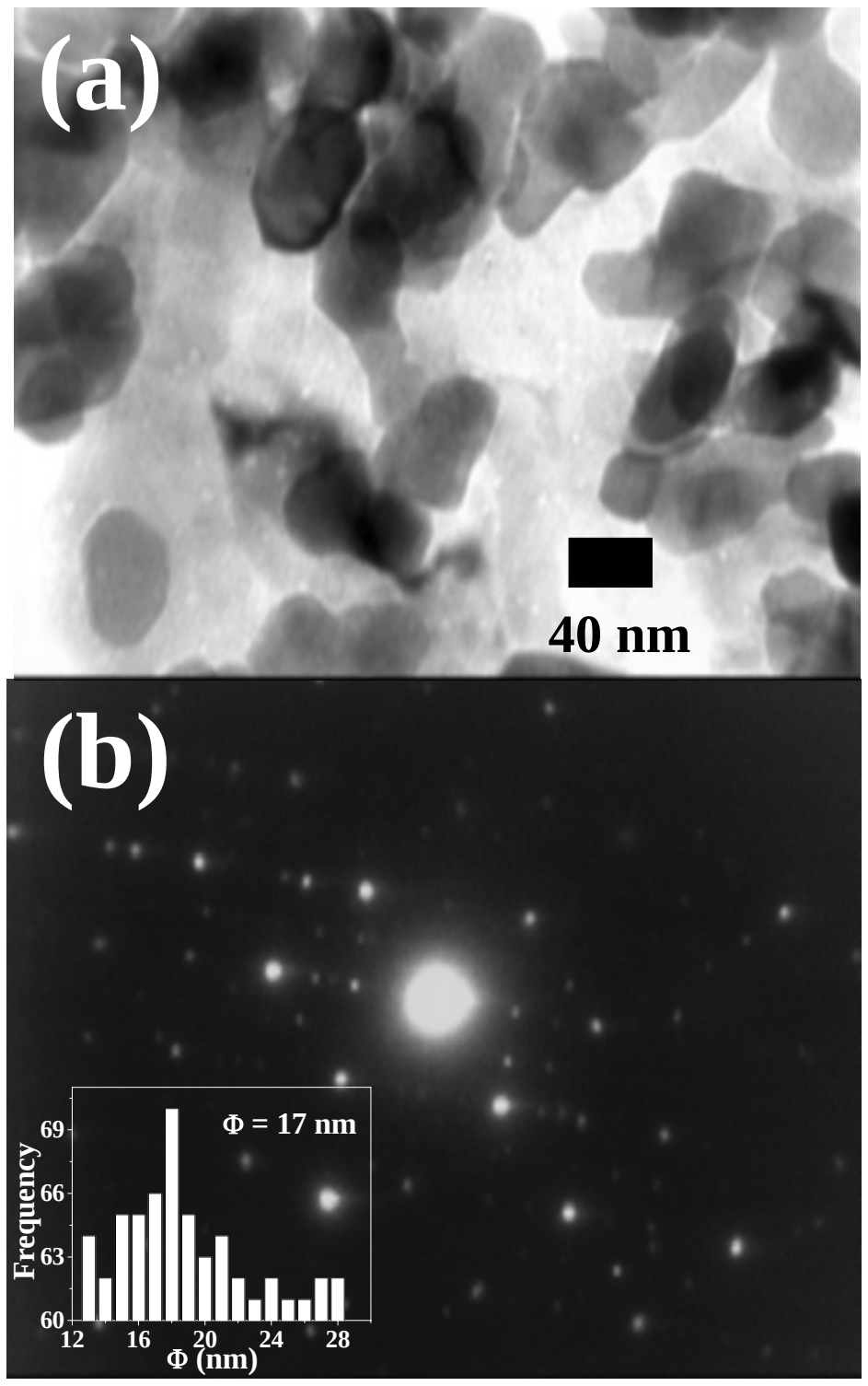

# Fig. 2

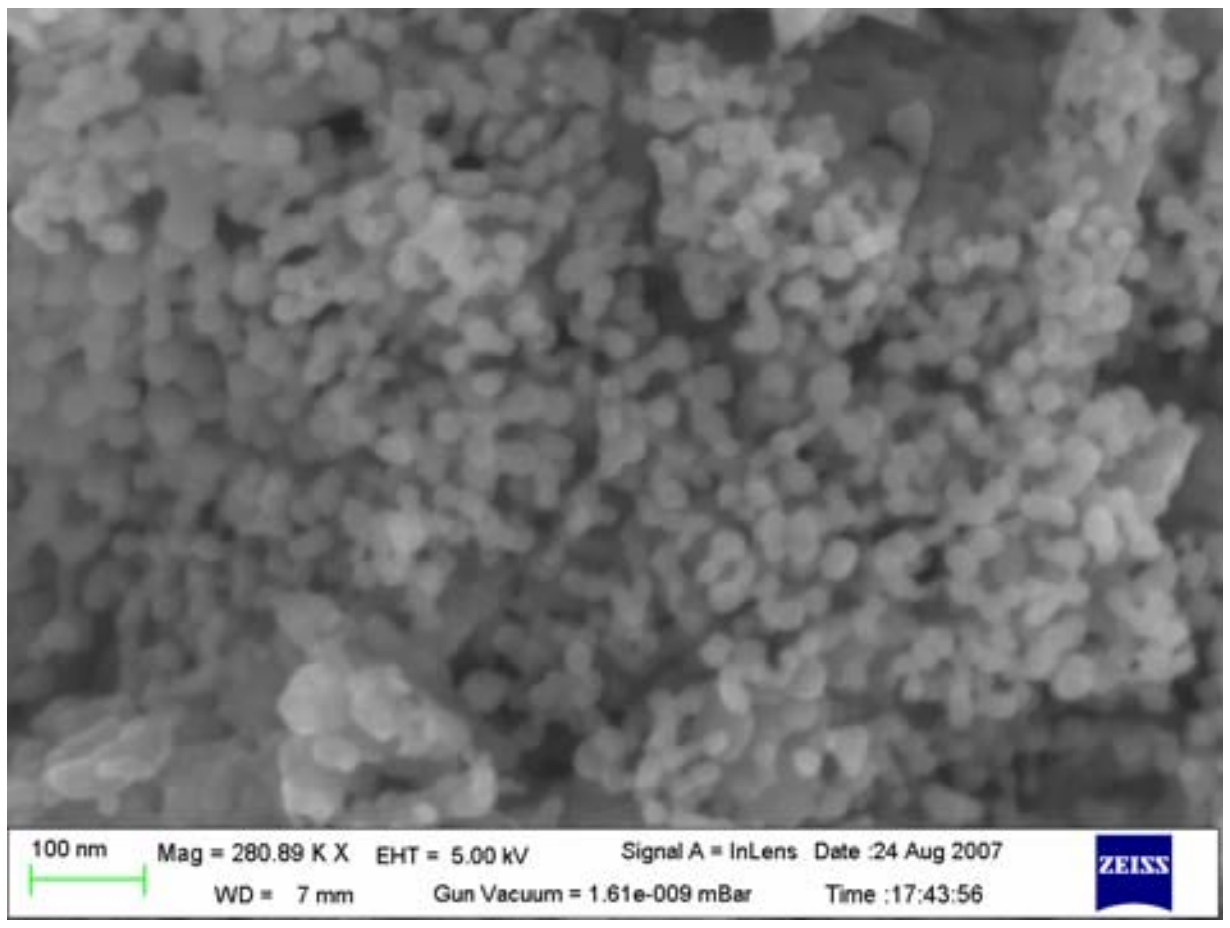
100 nm
Mag = 280.89 K X
EHT = 5.00 kV
Signal A = InLens
Date :24 Aug 2007
WD = 7 mm
Gun Vacuum = 1.61e-009 mBar
Time :17:43:56

# Fig. 3

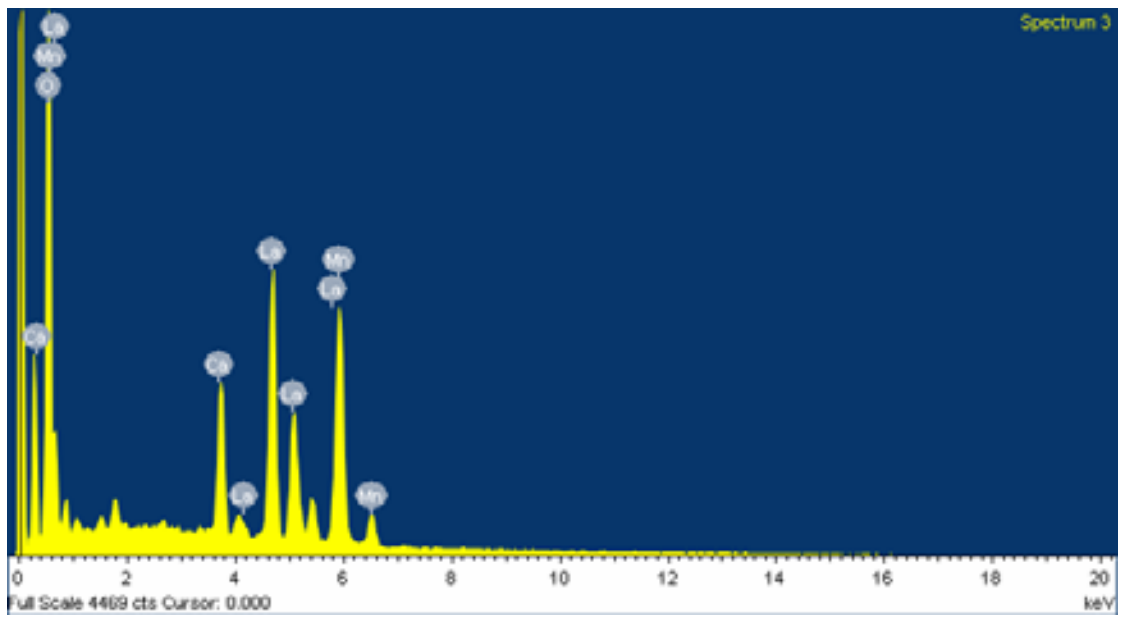

# Fig. 4

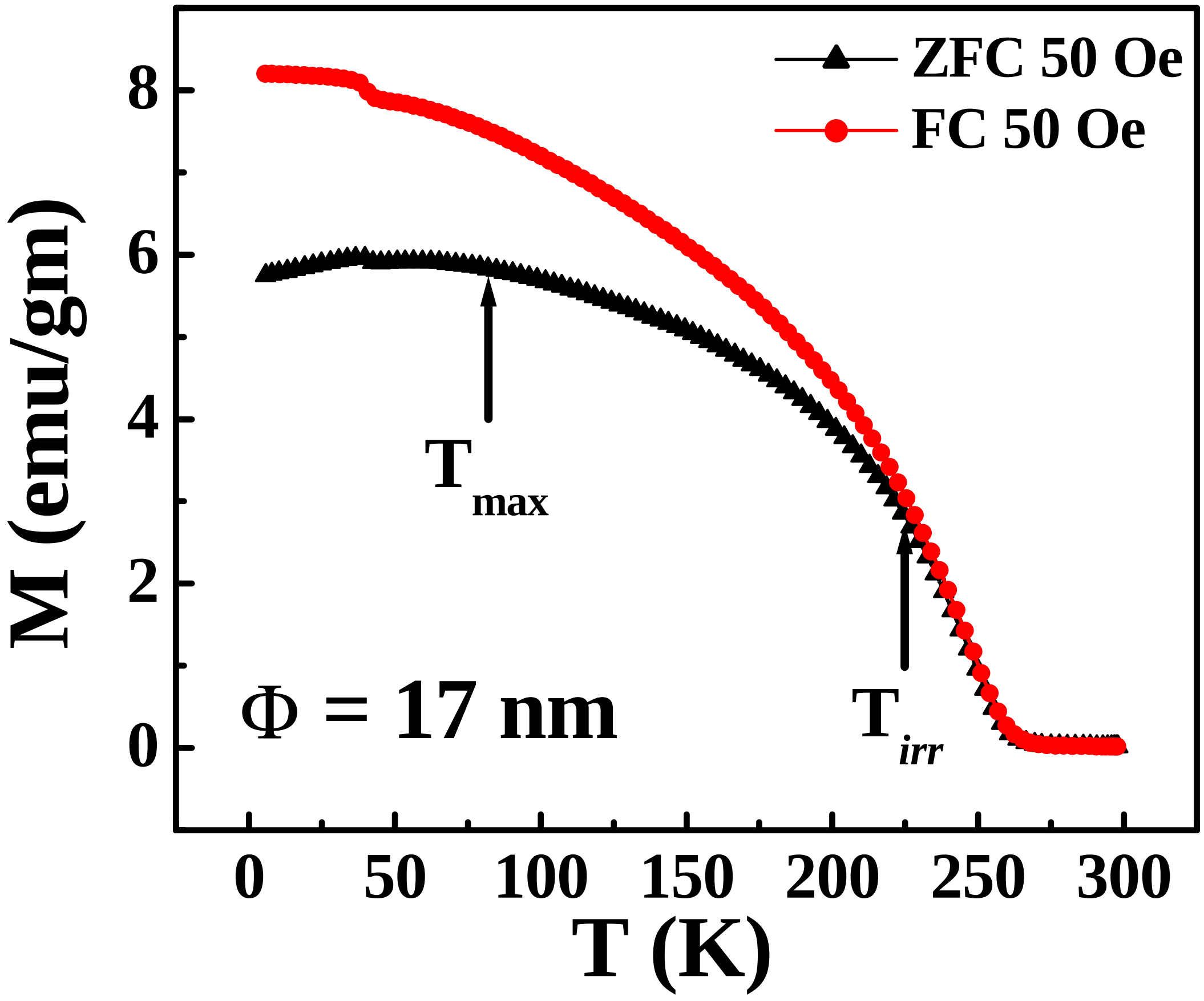

**Fig. 5**

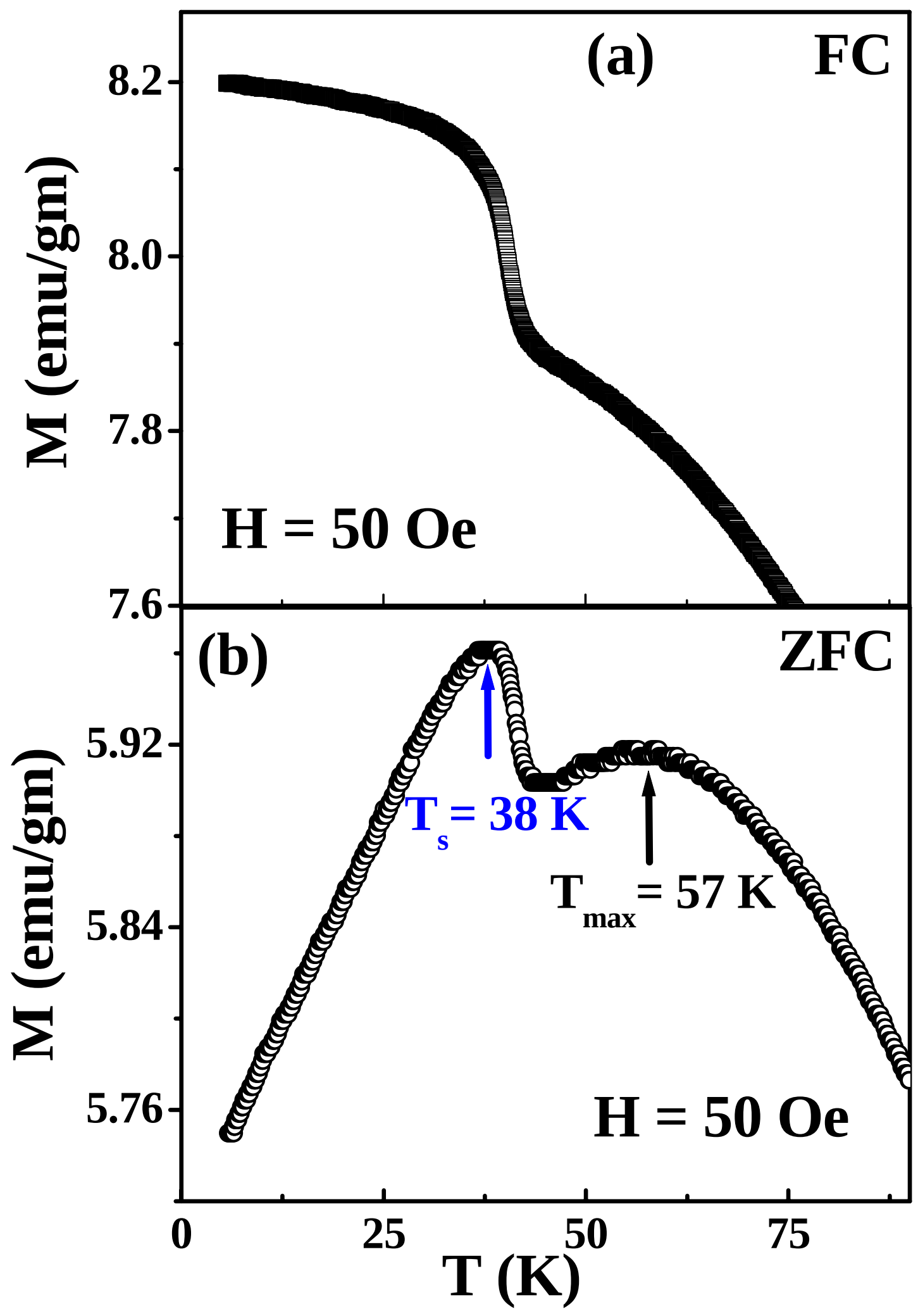

**Fig. 6**

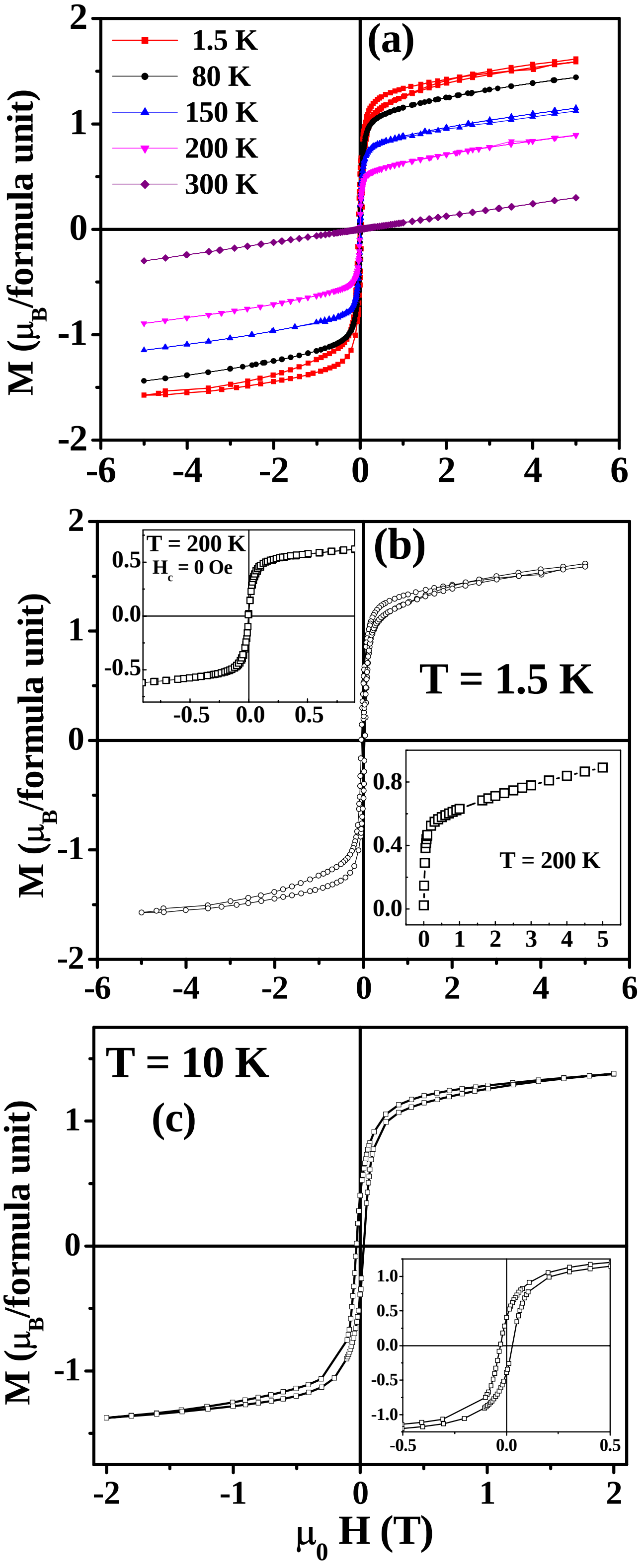

(a)
1.5 K
80 K
150 K
200 K
300 K
M (μB/formula unit)
(b)
T = 1.5 K
T = 200 K
Hc = 0 Oe
T = 200 K
(c)
T = 10 K
μ0 H (T)

# Fig. 7

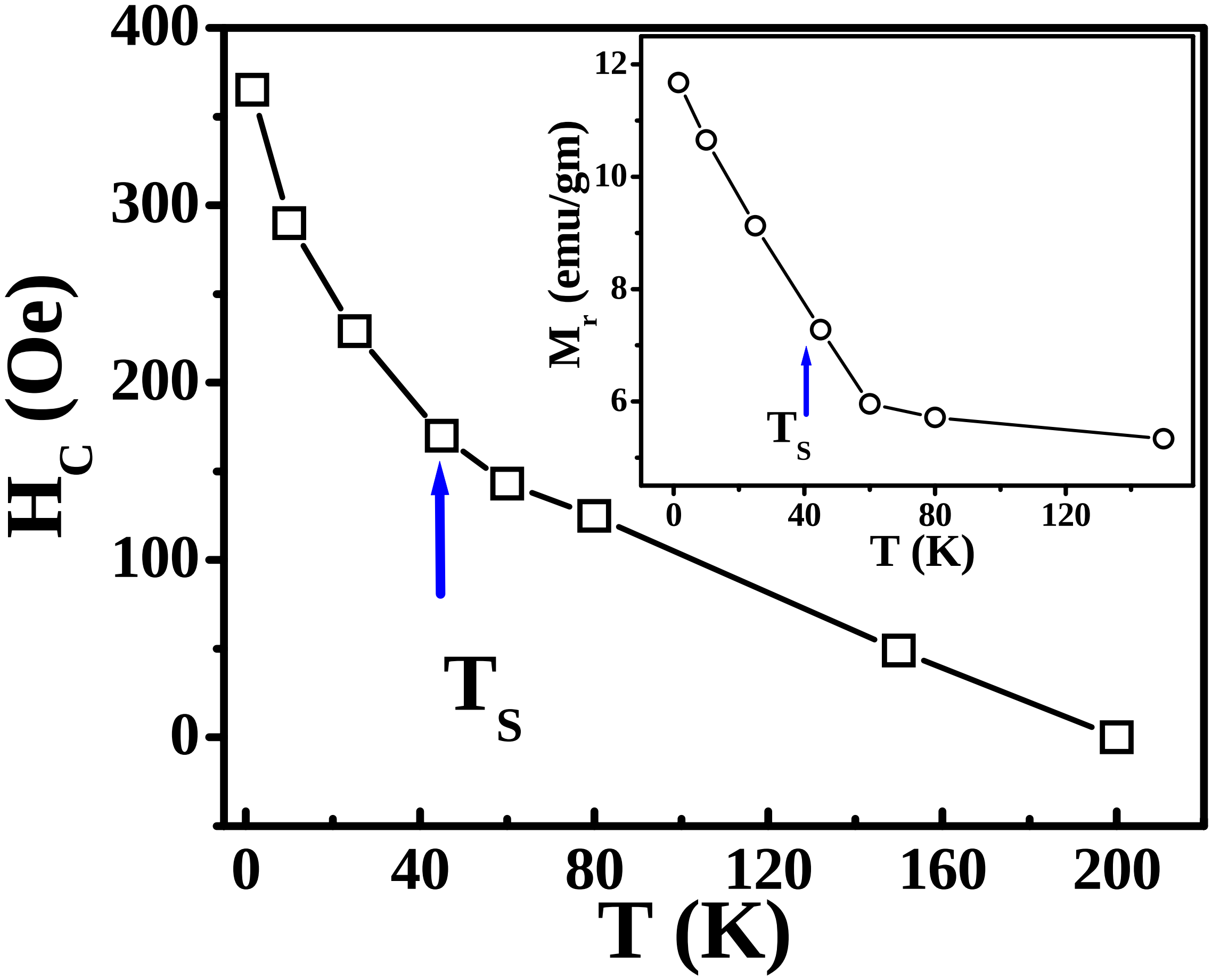

**Fig. 8**

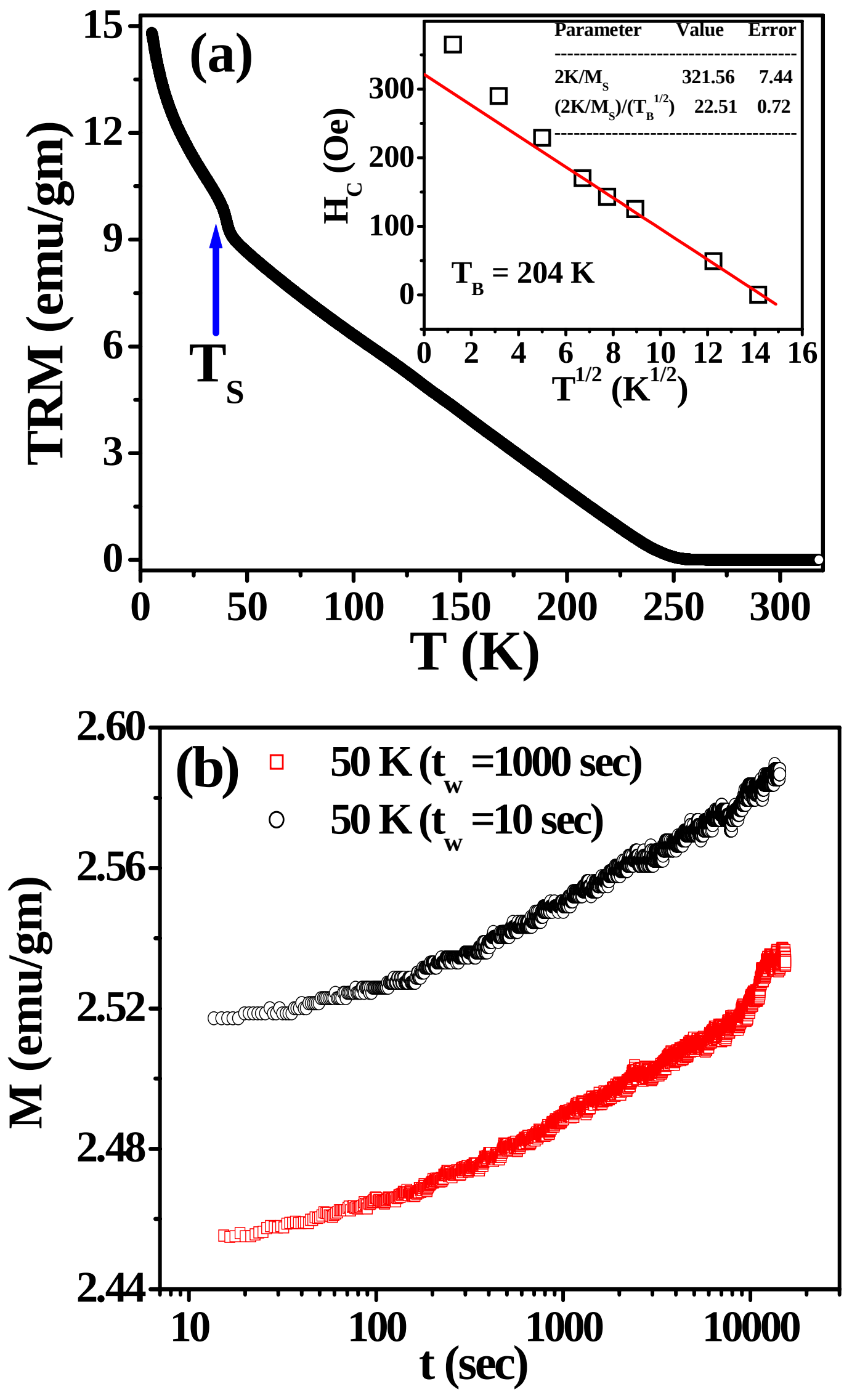

# Fig. 9

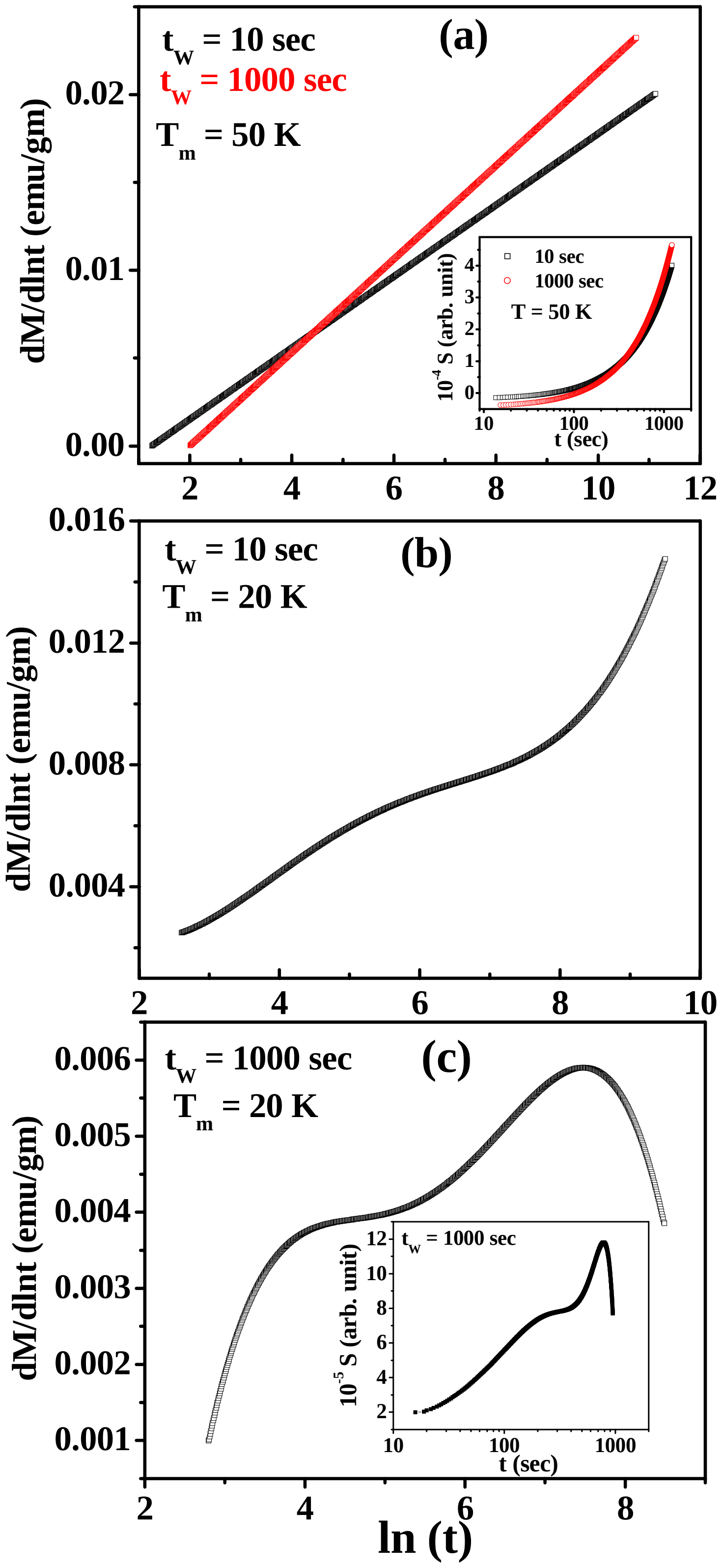
tW = 10 sec
tW = 1000 sec
Tm = 50 K
(a)
dM/dlnt (emu/gm)
0.00
0.01
0.02
2
4
6
8
10
12
10 sec
1000 sec
T = 50 K
10-4 S (arb. unit)
t (sec)
10
100
1000
tW = 10 sec
Tm = 20 K
(b)
0.016
0.012
0.008
0.004
tW = 1000 sec
Tm = 20 K
(c)
0.006
0.005
0.004
0.003
0.002
0.001
tW = 1000 sec
10-5 S (arb. unit)
ln (t)

# Fig. 10